\documentclass[pra,10pt,twocolumn,superscriptaddress,showpacs]{revtex4-2}

\usepackage{epsfig}
\usepackage{amsmath}
\usepackage{graphicx}
\usepackage{graphics}
\usepackage{float}
\usepackage{amssymb}
\usepackage{natbib}
\usepackage{epstopdf}
\usepackage{verbatim}
\usepackage{wrapfig}
\usepackage{mathtools}
\usepackage{amsfonts}
\usepackage{dsfont}
\usepackage{bm}
\usepackage{physics}

\usepackage{verbatim}

\renewcommand{\bra}[1]{\left\langle #1\right|}
\renewcommand{\ket}[1]{\left| #1\right\rangle}

\renewcommand{\tr}{\text{Tr}}
\newcommand{\tot}[1]{#1_{\text{tot}}}

\begin{document}

\title{A master equation incorporating the system-environment correlations present in the joint equilibrium state}

\author{Ali Raza Mirza}
\affiliation{School of Science \& Engineering, Lahore University of Management Sciences (LUMS), Opposite Sector U, D.H.A, Lahore 54792, Pakistan}

\author{Muhammad Zia}
\affiliation{School of Science \& Engineering, Lahore University of Management Sciences (LUMS), Opposite Sector U, D.H.A, Lahore 54792, Pakistan}

\author{Adam Zaman Chaudhry}
\email{adam.zaman@lums.edu.pk}
\affiliation{School of Science \& Engineering, Lahore University of Management Sciences (LUMS), Opposite Sector U, D.H.A, Lahore 54792, Pakistan}

\begin{abstract}
We present a general master equation, correct to second order in the system-environment coupling strength, that takes into account the initial system-environment correlations. We assume that the system and its environment are in a joint thermal equilibrium state, and thereafter a unitary operation is performed to prepare the desired initial system state, with the system Hamiltonian possibly changing thereafter as well. We show that the effect of the initial correlations shows up in the second-order master equation as an additional term, similar in form to the usual second-order term describing relaxation and decoherence in quantum systems. We apply this master equation to a generalization of the paradigmatic spin-boson model, namely a collection of two-level systems interacting with a common environment of harmonic oscillators, as well as a collection of two-level systems interacting with a common spin environment. We demonstrate that, in general, the initial system-environment correlations need to be accounted for in order to accurately obtain the system dynamics.
\end{abstract}


\maketitle

\section{Introduction}
The study of quantum systems interacting with their environment, namely open quantum systems, has gained immense importance lately. This is both due to practical reasons as well as the relevance of open quantum systems to the foundations of quantum mechanics \cite{BPbook}. In particular, precise quantum coherent control and state preparation necessary for quantum computation and information is only possibly by taking the interaction of a quantum system with its environment, while the process of decoherence sheds light on the so-called measurement problem. Amongst the variety of tools and techniques developed to tackle the dynamics of open quantum systems, the most popular approach is the quantum master equation approach whereby a differential equation for the time evolution of the system state is obtained and solved. The basic approach is that an initial state of the system and the environment is assumed, the total system-environment Hamiltonian is written down, and then using this total Hamiltonian, the time evolution of the total system-environment state is examined. Since we are typically interested only in the system dynamics, the environment is traced out to obtain the master equation. Unfortunately, performing this process to obtain the master equation for most realistic system-environment models involves making a series of approximations. For example, the system-environment coupling strength is assumed to be weak so that the joint time evolution operator can be found perturbatively \cite{BPbook,Weissbook}. The environment is assumed to have a short `memory time' (the Markovian approximation) - this means that the environment loses information about the system very quickly \cite{deVegaRMP,BreuerRMP}. Finally, the initial system-environment state is assumed to a simple product state, with the system and the environment independent of one another \cite{deVegaRMP}. 

With increasingly sophisticated quantum technologies, each of the assumptions typically made in the derivation of master equations has come under renewed reexamination. Master equations that allow one to deal with stronger system-environment coupling strengths have been formulated (see, for example, Ref.~\cite{MoussaPRA2014}). Measures of non-Markovianity have also been put forward \cite{BreuerRMP}. Most pertinent for us, the role of the initial system-environments - the correlations present in the total system-environment state at the initial state - has been investigated widely \cite{HakimPRA1985, HaakePRA1985, Grabert1988, SmithPRA1990, GrabertPRE1997, PazPRA1997, LutzPRA2003, BanerjeePRE2003, vanKampen2004, BanPRA2009, HanggiPRL2009, UchiyamaPRA2010, TanimuraPRL2010, SmirnePRA2010, DajkaPRA2010, ZhangPRA2010,TanPRA2011, CKLeePRE2012,MorozovPRA2012, SeminPRA2012,  ChaudhryPRA2013a,ChaudhryPRA2013b,ChaudhryCJC2013,FanchiniSciRep2014,FanSciRep2015correlation,ChenPRA2016,VegaRMP2017,VegaPRA2017,ShibataJPhysA2017,CaoPRA2017,MehwishEurPhysJD2019}. Such studies have generally been performed using exactly solvable models such as the pure dephasing model of a two-level system interacting with a collection of harmonic oscillators \cite{MorozovPRA2012,ChaudhryPRA2013a}. Some efforts have nevertheless been made to consider master equations that include the effect of the initial correlations. For example, in Ref.~\cite{ChaudhryPRA2013b}, the system and its environment were allowed to come to thermal equilibrium and thereafter a projective measurement was performed on the system to prepare the desired initial system state. It was shown that the effect of the initial correlations appears as an additional term in the second-order master equation, similar in form to the first term in the master equation that describes the effect of the free system-Hamiltonian. This approach was later generalized to higher order system-environment coupling strengths \cite{ShibataJPhysA2017}. Along similar lines, in this paper, we consider the quantum system and its environment to reach a joint equilibrium state. A unitary operator is then performed on the system to prepare (approximately) a desired initial system state, and a time-local master equation, correct to second-order in the system-environment coupling strength, that describes the ensuing dynamics of the system is derived. In fact, the system Hamiltonian before the application of the system Hamiltonian can be different from the system Hamiltonian after the unitary operator - the former plays a role in the initial state preparation, while the latter plays a role in the dynamics thereafter. We show that the effect of the initial correlations is now again contained in an additional term in the master equation, but now the form of this additional term is similar to that of the second term in the usual master equation that describes relaxation and decoherence of open quantum systems. We then apply our master equation to a collection of two-level systems interacting with a common environment of harmonic oscillators. We work out the additional term in the master equation, and perform numerical simulations to show that the effect of the initial correlations increases as the number of two-level systems increases. Along similar lines, we also apply our master equation to analyze the effect of initial correlations for a collection of two-level systems interacting with a spin environment.

This paper is organized as follows. In Sec.~II, we derive our general time-local second-order master equation. Sec.~III discusses the application of this master equation to the large spin-boson model, while Sec.~IV applies the master equation to a collection of two-level systems interacting with a spin environment. We then conclude in Sec.~V. The appendices consist of some technical details regarding the derivation of the usual relaxation term in the master equation, as well as the details of the exactly solvable limit of the large spin-boson model.

\section{The formalism}
We start by briefly discussing the problem we wish to solve. We are given a quantum system which is interacting with its environment and has reached a joint equilibrium state. At the initial time, a unitary operation is performed on the system alone, and the system Hamiltonian itself may also be changed. After all, one can, for example, apply a large magnetic field in order to prepare the initial state, and thereafter alter the value of this magnetic field to generate some desired system dynamics. Our problem is to derive the master equation, correct to second-order in the system-environment coupling strength, that describes the system dynamics. We then write the system-environment Hamiltonian as 
\begin{align}
    H_{tot} =
    \begin{cases}
      H_{S0} + H_B +\alpha V & t\le 0,\\
      H_{S} + H_B +\alpha V & t\ge 0.\\
    \end{cases}   
\end{align}
Here $H_S$ is the system Hamiltonian corresponding to coherent evolution of the system only after the initial time $t = 0$ at which the system state is prepared. $H_\text{S0}$ is similar to $H_S$
but with possibly different parameters in order to prepare the initial system state. $H_B$ is the Hamiltonian of the environment, and $V$ corresponds to the system-environment coupling. $\alpha$ is simply a dimensionless parameter introduced to keep track of the perturbation order; at the end of the calculation, we will set $\alpha = 1$. Let us now discuss in detail the initial state preparation.

\subsection{Initial state preparation}
We let our system to come to a joint equilibrium state with the environment. What we mean by this is that the equilibrium state of the system is not simply proportional to $e^{-\beta H_{S0}}$ - there are corrections due to the finite system-environment coupling strength. We instead consider the system and the environment together in the thermal equilibrium state proportional to  $e^{-\beta H_{\text{tot}}}$ with $H_{\text{tot}} =  H_{S0} + H_B +\alpha V$; the system state can then be obtained by simply tracing out the environment. A unitary operator $\Omega$ is then applied to the system. The initial system-environment state is then 
\begin{align}
    \rho_\text{tot}(0)
    &=\frac{\Omega e^{-\beta H_\text{tot}}\Omega^\dagger}{Z_\text{tot}}, \label{ref11}
\end{align}
with $Z_\text{tot}=\text{Tr}_\text{S,B}\left[e^{-\beta H_\text{tot}}\right]$ the partition function and $\tr_{S,B}$ denotes the trace over the system and the environment. Now, assuming the system-environment coupling strength to be weak, we can use the Kubo identity to expand the joint state given by Eq.~\eqref{ref11}. The Kubo identity tells us that for any two arbitrary operators $X$ and $Y$,
\begin{align}
    e^{\beta(X+Y)}&=e^{\beta X}\left[1+\alpha\int_{0}^{\beta} e^{-\lambda X}Y^{\lambda(X+Y)}d\lambda\right]. \label{ref12}
\end{align}
By setting $X=-(H_\text{S0}+H_B)$  and  $Y=-V$, and using the Kubo identity twice, we find that to second order in the system-environment coupling strength
\begin{align}
    &e^{-\beta(H_\text{S0}+H_B+V)}=e^{-\beta(H_\text{S0}+H_B)}\nonumber \\& -\alpha e^{-\beta(H_\text{S0}+H_B)}\int_{0}^{\beta} e^{\lambda(H_\text{S0}+H_B)}V  
    \times e^{-\lambda(H_\text{S0}+H_B)}d\lambda \nonumber \\ &+\alpha^2e^{-\beta(H_\text{S0}+H_B)}\int_{0}^{\beta}d\lambda e^{\lambda(H_\text{S0}+H_B)}V  \nonumber \\
    &\times e^{-\lambda(H_\text{S0}+H_B)}\int_{0}^{\lambda} e^{\lambda'(H_\text{S0}+H_B)} Ve^{-\lambda'(H_\text{S0}+H_B)}d\lambda'.\label{ref14}
\end{align}
We now write the system environment coupling $V$ as $F \otimes B$, where $F$ and $B$ are operators living in the system and environment Hilbert space respectively. The extension to the more general case where $V = \sum_\alpha F_\alpha \otimes B_\alpha$ is straightforward. Eq.~\eqref{ref14} can then be simplified as
\begin{align} 
    &e^{-\beta(H_\text{S0}+H_B+V)}=
    e^{-\beta(H_\text{S0}+H_B)} -\alpha e^{-\beta(H_\text{S0}+H_B)}\nonumber \\ &\times \int_{0}^{\beta} F(\lambda)\otimes B(\lambda)d\lambda + \alpha^2 e^{-\beta(H_\text{S0}+H_B)}\nonumber\\
   &\times \int_{0}^{\beta}d\lambda F(\lambda)\otimes B(\lambda)\int_{0}^{\lambda} F(\lambda')\otimes B(\lambda')d\lambda'.
\end{align}
where $F\left(\lambda\right)=e^{\lambda H_\text{S0}}Fe^{-\lambda H_\text{S0}}$ and $B\left(\lambda\right)=e^{\lambda H_B}Be^{-\lambda H_B}$.
We now use this in Eq.~\eqref{ref11} and thereafter take the trace over the environment to find the initial system state correct to second order in the system-environment coupling strength. This is important because our aim is to derive a master equation correct to second-order in the system-environment strength. For consistency, the initial system state used to solve this master equation should also be accurate to second-order in the system-environment coupling strength. For ease of notation, we write the initial system state as
\begin{align}
    \rho(0)=\rho^{(0)}(0)+\rho^{(1)}(0)+\rho^{(2)}(0),
    \label{sysdminitial}
\end{align}
where,
\begin{align}
    &\rho^{(0)}(0)
    =\frac{\text{Tr}_B\left[\Omega\left( e^{-\beta(H_\text{S0}+H_B)}\right)\Omega^\dagger\right]}{Z_\text{tot}}\\
    &\rho^{(1)}(0) \nonumber \\
    &=\frac{\text{Tr}_B\left[-\alpha \Omega\left(e^{-\beta(H_\text{S0}+H_B)}\int_{0}^{\beta} F(\lambda)\otimes B(\lambda)d\lambda\right)\Omega^\dagger\right]}{Z_\text{tot}}\\
    &\rho^{(2)}(0)= \frac{1}{{Z_\text{tot}}}\times \text{Tr}_B\Bigg[\alpha^2 \Omega\Big(e^{-\beta(H_\text{S0}+H_B)}\nonumber \\
    & \times \int_{0}^{\beta}d\lambda F(\lambda)\otimes B(\lambda) \int_{0}^{\lambda}F(\lambda')\otimes B(\lambda')d\lambda'\Big)\Omega^\dagger\Bigg]. 
\end{align}
Let us simplify these relations one by one. $\rho^{(0)}(0)$ can be simplified as,
\begin{align*}
   \rho^{(0)}(0)=\frac{e^{-\beta \widetilde{H}_\text{S0}} Z_B}{Z_\text{tot}},
\end{align*}
where $\widetilde{H}_{\text{S0}} = \Omega H_{\text{S0}}\Omega^\dagger$ and $Z_B= \text{Tr}_B\left[ e^{-\beta H_B}\right]$. As for $\rho^{(1)}(0)$, we can write
\begin{align*}
    \rho^{(1)}(0)= \frac{-\alpha Z_B\int_{o}^{\beta}\Omega e^{-\beta H_\text{S0}}F(\lambda)\Omega^{\dagger} \big\langle B(\lambda)\big\rangle_Bd\lambda}{Z_\text{tot}},
\end{align*}
where $\langle \hdots \rangle_B = \tr_B [e^{-\beta H_B} (\hdots)/Z_B]$. Since $\langle B(\lambda)\rangle_B$ is zero for most system-environment models, we simply get that $\rho^{(1)}(0)=0$. Carrying on, $ \rho^{(2)}(0)$ can be simplified as
\begin{align*}
    &\rho^{(2)}(0) = \nonumber\\
    &\frac{\alpha^2 Z_B\Omega e^{-\beta H_\text{S0}}\int_{0}^{\beta}\int_{0}^{\lambda} F(\lambda) F(\lambda') \Omega^{\dagger} \big\langle B(\lambda)  B(\lambda')\big\rangle_Bd\lambda'd\lambda}{Z_\text{tot}}. 
\end{align*}
To proceed further, we evaluate the partition function $Z_{\text{tot}}$. We note that $Z_{\text{tot}}$ has to be such that the trace of the system state $\rho(0)$ in Eq.~\eqref{sysdminitial} is one. It is then clear that 
\begin{align}
    Z_\text{tot}
    =&Z_B\text{Tr}_\text{S}\left[e^{-\beta H_\text{S0}} \right] +\alpha^2Z_B\text{Tr}_\text{S}\Bigg[\Omega e^{-\beta H_\text{S0}}  \nonumber \\
    &\times\int_{0}^{\beta}\int_{0}^{\lambda} F(\lambda) F(\lambda') \Omega^{\dagger} \big\langle B(\lambda)  B(\lambda')\big\rangle_Bd\lambda'd\lambda\Bigg]\notag.
\end{align}
It is then clear that the initial system density matrix can be written as 
\begin{align}
    \rho(0)
    &=\frac{e^{-\beta \widetilde{H}_{S0}}}{Z_{S0}Z'}\Big[\mathds{1}+\int_{0}^{\beta}\int_{0}^{\lambda} F^R(\lambda)F^R(\lambda')  \nonumber \\ 
    &\times\big\langle B(\lambda)  B(\lambda')\big\rangle_Bd\lambda'd\lambda\Big]
\end{align}
where $F^R(\lambda) = \Omega F(\lambda)\Omega^\dagger$, $Z_{S0} = \tr_S[e^{-\beta H_{S0}}]$, and 
\begin{align}
    Z'=1 + \int_{0}^{\beta}\int_{0}^{\lambda} \langle F(\lambda) F(\lambda') \rangle_S  
   \langle B(\lambda)  B(\lambda')\rangle_B d\lambda'd\lambda,
\end{align}
with $\langle \hdots \rangle_S = \tr_S [ e^{-\beta H_{S0}}(\hdots)/Z_{S0}]$. With the initial system state, correct to second order in the system-environment coupling strength available, we now turn our attention to deriving the second-order master equation.

\subsection{Derivation of the master equation}
We now derive a master equation that describes the time evolution of the system for $t > 0$. The system-environment Hamiltonian is 
\begin{align*}
    H_\text{tot} = H_\text{S}+H_\text{B}+\alpha V \equiv H_0 + \alpha V.
\end{align*}
Note that the system Hamiltonian $H_S$ can be different from the previous system Hamiltonian $H_{S0}$. Using perturbation theory, the unitary time evolution for such an Hamiltonian can be written as,
\begin{align}
    U (t) \approx U_0\left(t\right) \left[ 1 - \alpha \int_0^t U_0^\dagger (s) V U_0(s)\,ds \right] \label{ref15}
\end{align}
where, $U_0(t) \equiv U_S(t) \otimes U_B(t)$ is the free unitary time evolution corresponding to $H_0.$ The matrix elements of the system density matrix can be written as $\rho_{mn}\left(t\right) = \text{Tr}_S\left[\ket{n}\bra{m} \rho \left(t\right) \right]$, where $\ket{n}$ and $\ket{m}$ are some basis states of the system. Since $\rho(t) = \tr_B \rho_{\text{tot}}(t)$, we can alternatively write
\begin{align*}
    \rho_{mn}(t) = \text{Tr}_\text{S,B}\left[X_{nm}^H\left(t\right) \rho_\text{tot} (0) \right],
\end{align*}
where $X_{nm}^H(t) = U^\dagger(t) (\ket{n}\bra{m} \otimes \mathds{1}_B)U(t)$. The master equation can then be written in the general form 
\begin{align}
    \frac{d}{dt}\rho_{mn}(t) &=\text{Tr}_\text{S,B}\left[\rho_\text{tot}(0)\frac{d}{dt}X_{nm}^H(t)\right].
    \label{mastereqgen}
\end{align}
To make further progress, we note that $X_{nm}^H(t)$ is a Heisenberg picture operator. Using the Heisenberg equation of motion and Eq.~\eqref{ref15}, it can be shown that, correct to second order in the system-environment coupling strength, 
\begin{align} \label{ref16}
    \frac{d}{dt}X_{nm}^H(t) &=i\left[H_0^H(t),X_{nm}^H(t)\right]+i\alpha\left[\Tilde{V}(t),\Tilde{X}_{nm}(t)\right] \nonumber \\
    &+ \alpha^2\int_{0}^{t}ds\left[\left[\Tilde{V}(t),\Tilde{X}_{nm}(t)\right],\Tilde{V}(s)\right],
\end{align}
where the tildes denote time evolution under the `free' unitary operator $U_0 \left(t\right)$ while the superscript `$H$' denotes time evolution with the full time evolution operator. Using Eq.~\eqref{ref16} and given the initial system-environment state $\rho_\text{tot}(0)$ in Eq.~\eqref{sysdminitial}, we can derive our master equation that describes the time evolution of the quantum system by simplifying Eq.~\eqref{mastereqgen}. The first term is very straightforward. We simply have that 
\begin{align}
    &\text{Tr}_\text{S,B}\left[\rho_\text{tot}(0)i\left[H_0^H(t),X_{nm}^H(t)\right]\right] \nonumber\\
    &=i\text{Tr}_\text{S,B}\left[\rho_\text{tot}(t)\left[H_S + H_B,(\ket{n}\bra{m}\otimes \mathds{1}_B)\right]\right]\nonumber\\
    &=i\text{Tr}_\text{S}\left[\rho(t)\left[H_\text{S},\ket{n}\bra{m}\right]\right]\nonumber\\
    &=i\bra{m}\left[\rho(t),H_\text{S}\right]\ket{n}. \label{ref20}
\end{align}
This term simply tells us about free system evolution corresponding to the system Hamiltonian $H_\text{S}$.

To calculate the next term in our master equation, that is,
 \begin{align*}
    i\alpha \text{Tr}_\text{S,B}\left[\rho_\text{tot}(0)\left[V(t),X_{nm}(t)\right]\right],
\end{align*}
it is useful to write the initial system-environment state as $\rho_{\text{tot}}(0) = \rho_{\text{tot}}^{(0)} + \rho_{\text{tot}}^{(1)}$, where, using Eq.~\eqref{ref14}, it should be clear that 
\begin{align}
    \rho_\text{tot}^{(0)}(0)
    &=\frac{\Omega e^{-\beta(H_\text{S0}+H_B)}\Omega^\dagger}{Z_\text{tot}}  = \widetilde{\rho}_{S0} \otimes \rho_B \label{ref18}\\
    \rho_\text{tot}^{(1)}(0)
    &=\frac{-\alpha \Omega e^{-\beta(H_\text{S0}+H_B)}Q_\text{SB}(\beta)\Omega^\dagger}{Z_\text{tot}}, \label{ref19}
\end{align}
with $\widetilde{\rho}_{S0} = e^{-\beta \widetilde{H}_{S0}}/Z_{S0}$, $\rho_B = e^{-\beta H_B}/Z_B$, the partition function $\tot{Z} = Z_{S0} Z_B$ and $Q_{\text{SB}}(\beta) = \int_0^\beta d\lambda F(\lambda)\otimes B(\lambda)$. We do not need the higher order terms since there is already a factor of $\alpha$ in $i\alpha\left[V(t),X_{nm}(t)\right]$. Now, the contribution of $\rho_\text{tot}^{(0)}$ is
\begin{align*}
    &i\alpha \text{Tr}_\text{S,B}\left [\rho_\text{tot}^{(0)}(0)\left[U_0^\dagger(t)VU_0(t),U_0^\dagger(t)X_{nm}U_0(t)\right]\right]\\
    &=i\alpha \text{Tr}_\text{S,B}\left[\widetilde{\rho}_{S0}\otimes\rho_B U_0^\dagger(t) \left[F\otimes B,\ket{n}\bra{m}\otimes \mathds{1}_B\right] U_0(t)\right]\\
    &=i\alpha \text{Tr}_\text{S}\left [\Tilde{\rho}_{S0}U^\dagger(t)\left[F,Y_{nm}\right]U(t)\right ]\times \langle B(t)\rangle_B.
\end{align*}
Again, since $\langle B(t)\rangle_B$ is usually zero for most system-environment models, this contribution turns out to be zero. 

The most interesting contribution is due to $\rho_\text{tot}^{(1)}(0)$. This is
\begin{align}
    &i\alpha \text{Tr}_\text{S,B}\Big[\rho_\text{tot}^{(1)}(0)\left[U_0^\dagger(t)VU_0(t),U_0^\dagger(t)X_{nm}U_0(t)\right]\Big]\nonumber\\
    &=\frac{-i\alpha^2}{Z_\text{S0}}\int_{0}^{\beta}\text{Tr}_\text{S,B}\Big[\rho_B{\Omega e^{-\beta H_\text{S0}} F(\lambda)\Omega^\dagger\otimes  B(\lambda)}U_S^\dagger(t)  \nonumber \\
    & \times\left[F,\ket{n}\bra{m}\right]U_S(t)U_B^\dagger(t)BU_B(t)\Big]d\lambda\nonumber\\
    &=\frac{-i\alpha^2}{Z_\text{S0}}\int_{0}^{\beta}\text{Tr}_\text{S}\Big[{\Omega e^{-\beta H_\text{S0}} F(\lambda)\Omega^\dagger}U_S^\dagger(t)\left[F,\ket{n}\bra{m}\right] \nonumber \\
    &\times U_S(t)\Big] \text{Tr}_B\Big[\rho_B B(\lambda)B(t)\Big]d\lambda\nonumber\\
    &=\frac{-i\alpha^2}{Z_\text{S0}}\int_{0}^{\beta}\bra{m}\left[U_S(t)\Omega e^{-\beta H_\text{S0}} F(\lambda)\Omega^\dagger U_S^\dagger(t),F\right]\ket{n}  \nonumber \\
    & \times B_{\text{corr}}(\lambda,t)\, d\lambda, \label{ref21}.
\end{align}
where we have defined $B(t) = U_B^\dagger (t) B U_B(t)$ and $B_{\text{corr}}(\lambda,t) = \tr_B \Big[\rho_B B(\lambda)B(t)\Big]$. This is additional term in the master equation that takes into account the effect of the initial correlations, correct to second order in the system-environment coupling strength. In basis-independent form, we can write this term as 
\begin{align}
    -i\left[ \widetilde{\rho}(t)J^R_\text{corr}(\beta,t),F\right],
\end{align}
where we have defined 
\begin{align} \label{ref23}
    J^R_\text{corr}(\beta,t)
    &=\int_{0}^{\beta}  \overleftarrow{F}^R(\lambda,t)\text{B}_\text{corr}(\lambda,t)d\lambda,\\
    \overleftarrow{F}^R(\lambda,t)
    &= U_S(t)\Omega e^{\lambda H_\text{S0}}Fe^{-\lambda H_\text{S0}}\Omega^\dagger U_S^\dagger(t).
\end{align}
We now write 
\begin{align}
    -i\left[ \widetilde{\rho}(t)J^R_\text{corr}(\beta,t),F\right] = -\frac{i}{2}\left(\left[ \widetilde{\rho}(t)J^R_\text{corr}(\beta,t),F\right] - \text{h.c.}\right).
\end{align}
This is permitted because $ -i\left[ \widetilde{\rho}(t)J^R_\text{corr}(\beta,t),F\right]$ is hermitian, so 
$\left[ \widetilde{\rho}(t)J^R_\text{corr}(\beta,t),F\right]$ is anti-hermitian. We now replace $\widetilde{\rho}(t)$ by $\rho(t)$. This step is permitted since the corrections are of order higher than the second-order master equation that we are considering. Consequently, the term in the master equation that takes into account the initial correlations is $-\frac{i}{2}\left(\left[ \rho(t)J^R_\text{corr}(\beta,t),F\right] - \text{h.c.}\right)$.

	We next simplify the contribution of the third term in Eq.~\eqref{ref16}. It is clear that now only $\rho_{\text{tot}}^{(0)}(0)$ contributes. Similar manipulations to those performed above lead to (see the appendix for details) 
$$ \alpha^2\int_{0}^{t}\bra{m}\Big(\left[\Bar{F}(t,s)\widetilde{\rho}_{S}(t),F\right]C_{ts}+\text{h.~c.}\Big)\ket{n}\,ds, $$
where the environment correlation function is $C_{ts} = \langle B(t)B(s) \rangle_B$, $\widetilde{\rho}_S(t) = U_S(t) \widetilde{\rho}_{S0} U_S^\dagger (t)$, $\Bar{F}(t,s)=U_S^\dagger(t,s)FU_S(t,s)$, and $\text{h.~c.}$ denotes the hermitian conjugate. We can further replace $\widetilde{\rho}_S(t)$ by $\rho(t)$ to get 
$$ \alpha^2\int_{0}^{t}\bra{m}\Big(\left[\Bar{F}(t,s)\rho(t),F\right]C_{ts}+\text{h.~c.}\Big)\ket{n}\,ds. $$
Once again, this is permitted since the corrections lead to terms of higher order in the master equation (compared to the second order master equation that we are considering). We now put all the terms together to arrive at the general basis-independent form of the master equation given by 

\begin{align}
    &\frac{d}{dt}\rho(t)
    =i\left[\rho(t),H_S\right]-\frac{i}{2}\left(\left[ \rho(t)J^R_\text{corr}(\beta,t),F\right] - \text{h.c.}\right)\nonumber\\
    &+\int_{0}^{t}\left(\left[\Bar{F}(t,s)\rho(t),F\right]C_{ts}+\text{h.c.}\right)\,ds.
     \label{finalme}
\end{align}

\section{Application to the large spin-boson model}
We will now apply our derived master equation to a variant of the paradigmatic spin-boson model \cite{BPbook} with a large number of two-level systems interacting with a common environment of harmonic oscillators \cite{KurizkiPRL20112,ChaudhryPRA2013a,ChaudhryPRA2013b}. Recall that the total system-environment Hamiltonian is given by $H_\text{tot} = H_\text{S0}+H_\text{B}+V$ for $t < 0$, while $H_{\text{tot}} = H_S+H_B+V$ for $t \geq 0$. For the large spin-boson model, we write
\begin{align}
    H_\text{S0}
    &= \varepsilon_0 J_z + \Delta_0 J_x\\
    H_S&= \varepsilon J_z + \Delta J_x\\
    H_B&= \sum_{k} \omega_k b_{k}^{\dagger} b_k,\\
    V&=J_z \sum_k \big( g_k^* b_k + g_k b_k^\dagger \big)
\end{align}
where $J_x,J_y,J_z$ are the collective spin operators with $J^2 =J_x^2+J_y^2+J_z^2, \varepsilon$ is energy bias, $\Delta$ is the tunneling amplitude, $H_B$ is the bath of harmonic oscillators (we are ignoring zero point energy), while $V$ describes the interaction between the common harmonic oscillator bath and the spin system. We have set $ \hbar =1 $ throughout and the values of other parameters will be in dimensionless units. Note that the system operator $F=J_z,$ and the bath operator $B=\sum_k \big( g_k^* b_k + g_k b_k^\dagger \big). $ One imagine that the large-spin system has been interacting with the environment for a long time with a relatively large value of $\varepsilon_0$ and a small value of $\Delta_0$. In such a situation, the state of the system will be approximately corresponding to the state with all spins down in the $z$-direction. At time $t = 0$, we then apply a unitary operator to prepare the desired initial state. For example, if the desired initial state is one with all spins in the $x$-direction. then the unitary operator that should be applied is $U_R = e^{i\pi J_y/2}$. In other words, a $\frac{\pi}{2}$-pulse is used to prepare the initial system state, with the assumption that this pulse takes a very short time to be applied. With the initial state approximately prepared, we can then change the parameters to the system Hamiltonian to whatever values we desire to generate any required system evolution. Let us look at how the initial system-environment correlations appear in this system evolution using our general master equation. 

\begin{figure}[htp] 
  \centering 
  \includegraphics[width=0.95\linewidth]{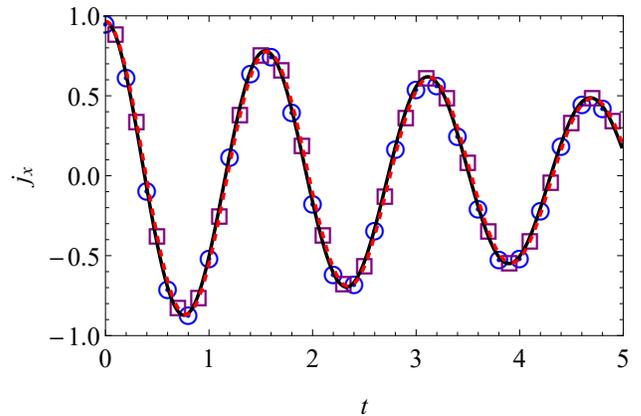}
\caption{Behavior of $j_x$ versus $t$ for N = 1 using the exact solution with (blue circled dot) and without (purple squares) initial correlations, as well as using the master
equation with (solid, black line) and without (dashed, red line)initial correlations. We have used $\varepsilon=\varepsilon_0=4$, $G = 0.05$, $\beta=1$ and $\omega_c=5$. Here and in all other figures, the plotted variables are all in dimensionless units} \label{Puredephasing-N=1-unitary}
\end{figure}

\begin{figure}[htp] 
  \centering 
  \includegraphics[width=0.95\linewidth]{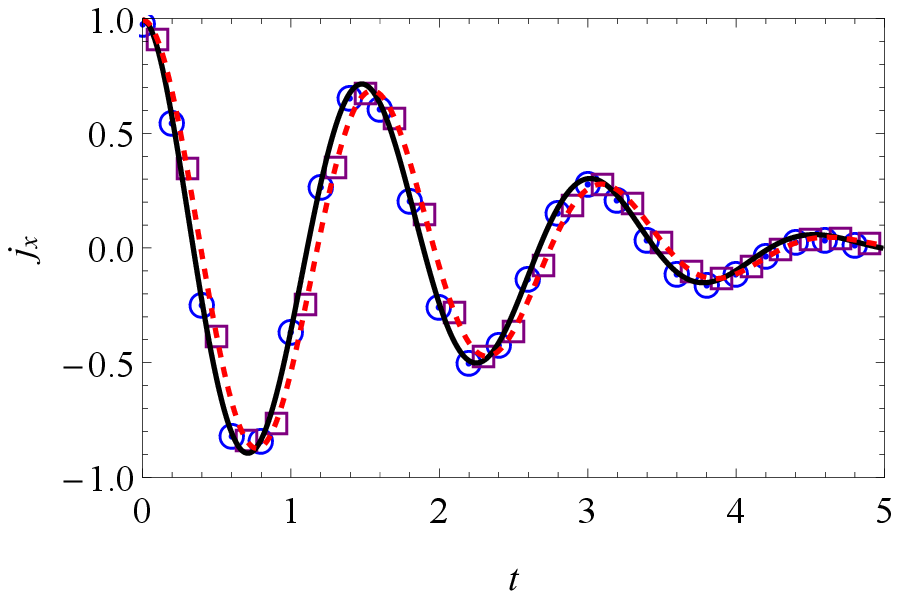}
\caption{Same as Fig. \ref{Puredephasing-N=1-unitary}, except that we now have $N = 4$.}
\label{Puredephasing-N=4-unitary}
\end{figure}

\begin{figure}[htp] 
  \centering 
  \includegraphics[width=0.95\linewidth]{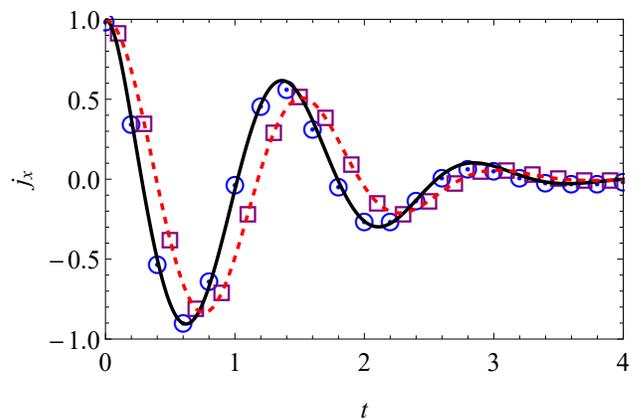}
\caption{Same as Fig. \ref{Puredephasing-N=4-unitary}, except that we now have $N = 10$.} \label{Puredephasing-N=10-unitary}
\end{figure}

Our objective is to calculate the operator $J^R_{\text{corr}}$. To do so, we first find 
\begin{align*}
    &\overleftarrow{F}^R(\lambda,t)=U(t)\left[U_R\left(e^{\lambda H_{S0}}Fe^{-\lambda H_{S0}}\right)U^\dagger_R\right]U^\dagger(t)\\
    &=J_x\left[a_xd_x+a_yc_x-a_z b_x\right]+J_y\left[a_xd_y+a_yc_y-a_z b_y\right]\\
    &+J_z\left[a_xd_z+a_yc_z-a_zb_z\right],
\end{align*}
with 
\begin{align*}
       a_x
    &=\frac{\varepsilon_0\Delta_0}{\Delta'^2} \left\{1-\cosh\left({\lambda\Delta'} \right)\right\},\\
    a_y
    &=\frac{-i\Delta_0}{\Delta'}\sinh \left(\lambda\Delta'\right),\\
    a_z
    &= \frac{\varepsilon_0^2 + \Delta^2_0 \cosh \left({\lambda\Delta'}\right)}{\Delta'^2},\\
     b_x
    &= \frac{\Delta^2 + \varepsilon^2 \cos \left({\widetilde{\Delta}t}\right)}{\widetilde{\Delta}^2},\\
    b_y
    &= \frac{ \varepsilon}{\widetilde{\Delta}}\sin \left( \widetilde{\Delta}t\right),\\
    b_z
    &= \frac{\varepsilon \Delta}{\widetilde{\Delta}^2} \left\{ 1 - \cos\left({\widetilde{\Delta}t} \right)\right\},\\
    c_x 
    &= -\frac{\varepsilon }{\widetilde{\Delta}}\sin{\left(\widetilde{\Delta}t\right)},\\
    c_y
    &= \cos{\left(\widetilde{\Delta}t\right)},\\
    c_z
    &= \frac{\Delta }{\widetilde{\Delta}}\sin{\left(\widetilde{\Delta}t\right)},\\
     d_x &= \frac{\varepsilon\Delta }{\tilde{\Delta}^2}\left\{1-\cos{\left(\widetilde{\Delta}t\right)}\right\},\\
    d_y &=-\frac{\Delta }{\widetilde{\Delta}}\sin{\left(\widetilde{\Delta}t\right)},\\
    d_z &= 1+\frac{\Delta^2 }{\tilde{\Delta}^2}\left\{\cos{\left(\widetilde{\Delta}t\right)}-1\right\}.
\end{align*}
Here $\Delta'^2 = \varepsilon_0^2 + \Delta_0^2$ and $\widetilde{\Delta}^2 = \varepsilon^2 + \Delta^2$. In short, 
\begin{align} \label{ref24}
    \overleftarrow{F}^R(\lambda,t)=\alpha_1(\lambda,t)J_x+\alpha_2(\lambda,t)J_y+\alpha_3(\lambda,t)J_z,
\end{align}
where,
\begin{align*}
    \alpha_1(\lambda,t)
    &=a_xd_x+a_yc_x-a_z b_x,\\
    \alpha_2(\lambda,t)
    &=a_xd_y+a_yc_y-a_z b_y,\\
    \alpha_3(\lambda,t)
    &=a_xd_z+a_yc_z-a_zb_z.
\end{align*}
It then follows that 
\begin{align}
\label{initialcorrelations}
    J_\text{corr}^R(\beta,t)
    &=P(\beta,t)J_x+Q(\beta,t)J_y+R(\beta,t)J_z,
\end{align}
with,
\begin{align*}
    P(\beta,t)
    &=\int_{0}^{\beta}  \alpha_1(\lambda,t)B_{\text{corr}}(\lambda,t)d\lambda\\
    Q(\beta,t)
    &=\int_{0}^{\beta}  \alpha_2(\lambda,t)B_{\text{corr}}(\lambda,t)d\lambda\\
    R(\beta,t)
    &=\int_{0}^{\beta}  \alpha_3(\lambda,t)B_{\text{corr}}(\lambda,t)d\lambda.
\end{align*}
We now calculate bath correlation term $B_{\text{corr}}(\lambda,t)$. First, 
\begin{align}
    B(t)=\sum_k\left(g_k^*e^{-i\omega_{k}t}b_k+g_ke^{i\omega_{k}t}b_k^\dagger\right),
\end{align}
and,
\begin{align}
    B(\lambda)=\sum_k\left(g_k^*e^{-\lambda\omega_{k}}b_k+g_ke^{\lambda\omega_{k}}b_k^\dagger\right).
\end{align}
Since $B_{\text{corr}}(\lambda,t) =\text{Tr}[\rho_B B(\lambda) B(t)]$, we find 
\begin{align}
    B_{\text{corr}}(\lambda,t)&=\sum_k  |g_k|^2\Big\{e^{-\omega_{k}\left(\lambda-it\right)}\nonumber \\& +2n_k\cosh{\left(\lambda\omega_{k}-i\omega_{k}t\right)}\Big\},\label{ref26}
\end{align}
with $n_k$ given by Bose-Einstein statistics as 
\begin{align}
    n_k&=\frac{1}{2}\left\{\coth{\left(\frac{\beta\omega_k}{2}\right)}-1\right\}\nonumber.
\end{align}
To perform the sum over the environment modes, we use the spectral density $J(\omega)$ via $\sum_k |g_k|^2 (\hdots) \rightarrow \int_0^\infty d\omega \, J(\omega) (\hdots)$. We generally use an Ohmic spectral density of the form $J(\omega) = G\omega e^{-\omega/\omega_c}$. The integrals are performed numerically to find $J_{\text{corr}}(\beta,t)$, and the results are incorporated in the numerical simulations of the master equation. We start by first looking at the pure dephasing case ($\Delta=\Delta_0=0$), since this case can be solved exactly and serves as a useful benchmark (see the appendix for details regarding the exact solution). We illustrate our results in Fig.~\ref{Puredephasing-N=1-unitary} for $N = 1$. Two points should be noted. First, the role played by initial correlations is very small in this case. Second, our master equation reproduces the exact results very well. Since we expect that the role of the initial correlations increases with increasing $N$, we next look at $N = 4$ and $N = 10$. Results are shown in Figs.~\ref{Puredephasing-N=4-unitary} and \ref{Puredephasing-N=10-unitary}. It is clear that as $N$ increases, the initial correlations play a larger and larger role. Moreover, the extra term in our master equation is able to take into account the effect of the initial correlations very well.


\begin{figure}[htp] 
  \centering 
  \includegraphics[width=0.95\linewidth]{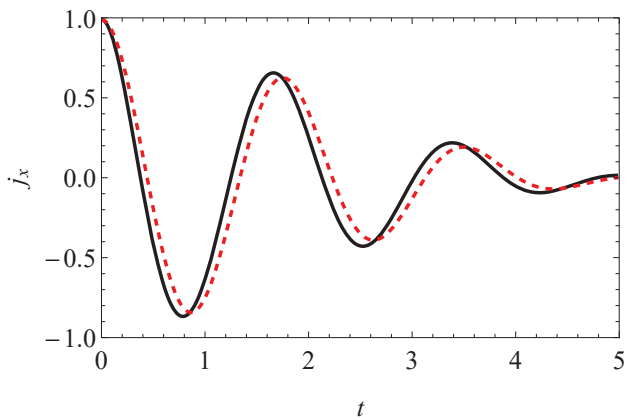}
\caption{Behavior of $j_x$ against $t$ for $N=2$ with (black, solid) and without (dashed, red) taking into
account initial correlations. Here we have used $\varepsilon_0 = 4$, $\varepsilon=2.5$ and $\Delta=\Delta_0=0.5$, while the rest of the parameters used are the same as Fig. \ref{Puredephasing-N=1-unitary}.} \label{Beyond-PD-N=2}
\end{figure}

\begin{figure}[htp] 
  \centering 
  \includegraphics[width=0.95\linewidth]{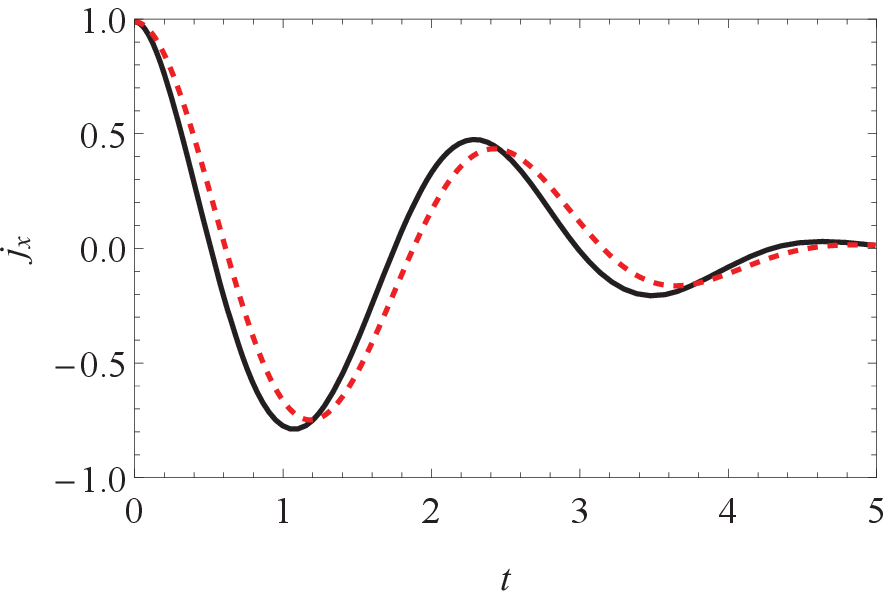}
\caption{Same as Fig. \ref{Beyond-PD-N=2}, except that we now have $N = 4$.} \label{Beyond-PD-N=4}
\end{figure}

\begin{figure}[htp] 
  \centering 
  \includegraphics[width=0.95\linewidth]{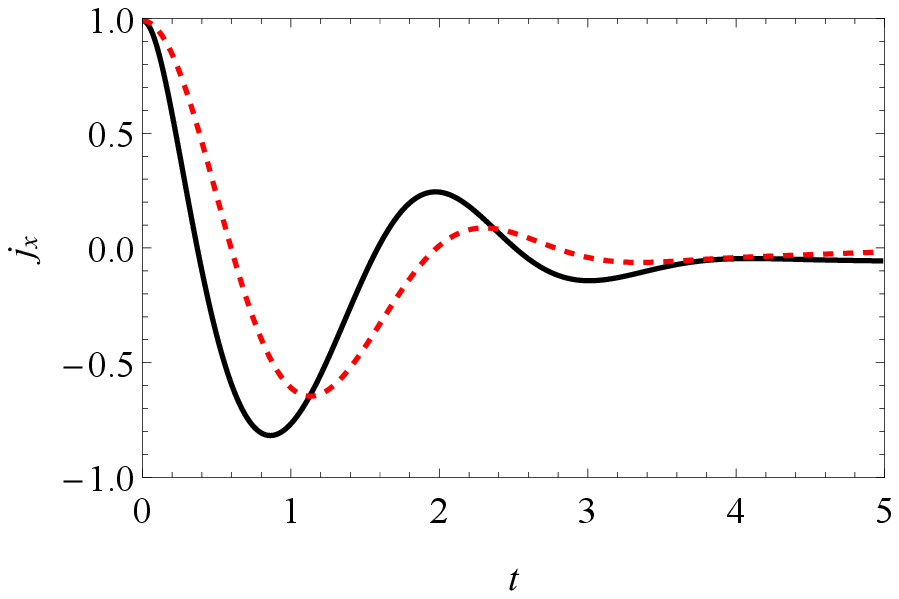}
\caption{Same as Fig. \ref{Beyond-PD-N=4}, except that we now have $N = 10$.} \label{Beyond-PD-N=10}
\end{figure}

\begin{figure}[htp] 
  \centering 
  \includegraphics[width=0.95\linewidth]{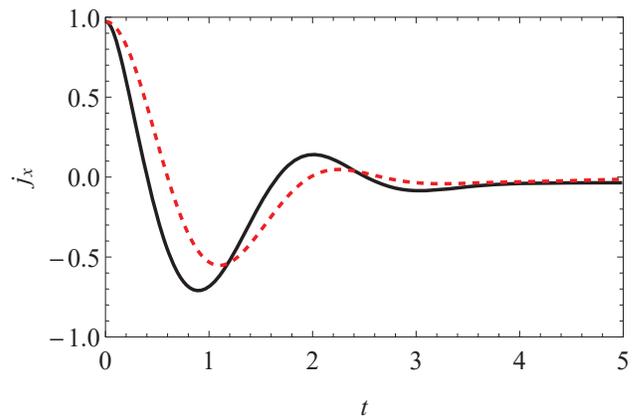}
\caption{Behavior of $j_x$ against $t$ for $N=10$ with (black, solid) and without (dashed, red) taking into account initial correlations. Here we have used the same parameters as Fig.~\ref{Beyond-PD-N=10}, except that $\beta = 0.5$.} \label{Beta=0.5}
\end{figure}

\begin{figure}[htp] 
  \centering 
  \includegraphics[width=0.95\linewidth]{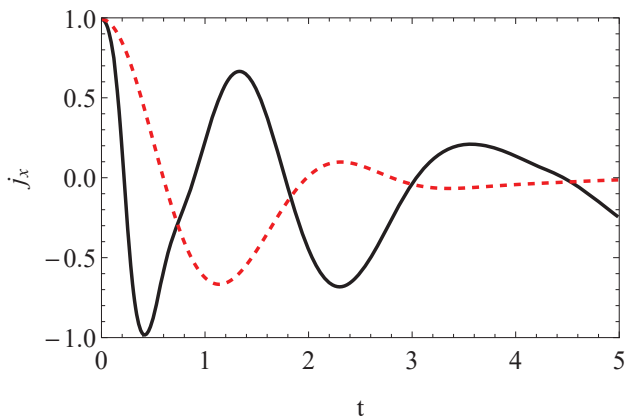}
\caption{Same as Figs.~\ref{Beyond-PD-N=10} and \ref{Beta=0.5}, except that we now have $\beta=1.5$.} \label{Beta=1.5}
\end{figure}

Having shown that our master equation is able to reproduce results for the pure dephasing model, we are now in a position to go beyond the pure dephasing model and see the effects of the initial correlations. In Fig.~\ref{Beyond-PD-N=2}, we have shown the dynamics of $j_x$ with a non-zero value of the tunneling amplitude for $N = 2$. It is clear that the initial correlations do have a small influence on the dynamics. This effect becomes more pronounced as we increase $N$ (see Figs.~\ref{Beyond-PD-N=4} and \ref{Beyond-PD-N=10}). We have also looked at how the role played by the initial correlations changes as the temperature changes. To this end, we compare Fig.~\ref{Beyond-PD-N=10}, where the inverse temperature is $\beta = 1$, with Fig.~\ref{Beta=0.5} where $\beta = 0.5$ and Fig.~\ref{Beta=1.5} where $\beta = 1.5$. As expected, at higher temperatures, the effect of the initial correlations decreases, while at lower temperatures, the effect of the initial correlations increases. \\


\begin{figure}[htp] 
  \centering 
  \includegraphics[width=0.95\linewidth]{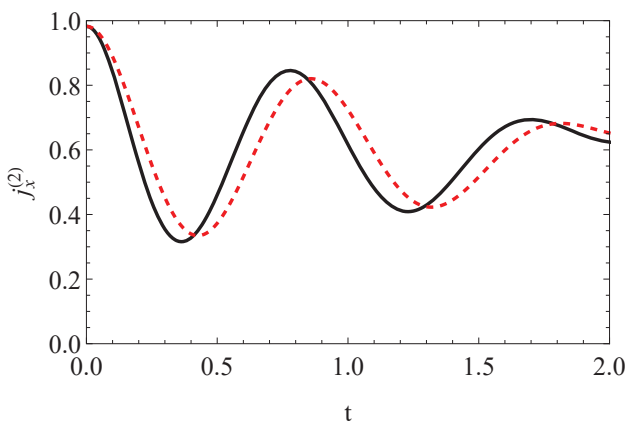}
\caption{Behavior of $j_x^{(2)}$ against $t$ for $N = 4$
with (black, solid) and without (dashed, red) taking into account initial correlations. The rest of the parameters used are the same as those in Fig.~\ref{Beyond-PD-N=2}.} \label{Jx^2-plot-for-N=4}
\end{figure}

\begin{figure}[htp] 
  \centering 
  \includegraphics[width=0.95\linewidth]{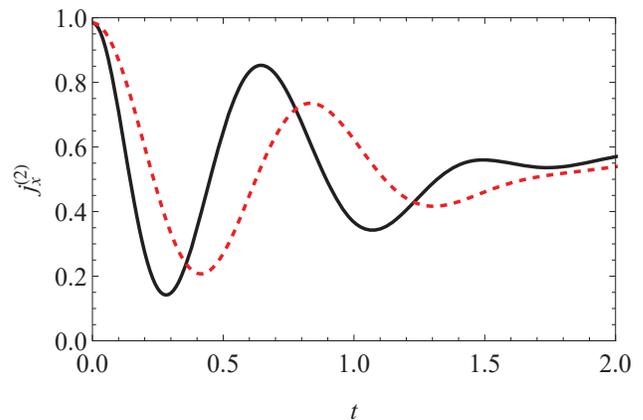}
\caption{Same as Fig. \ref{Jx^2-plot-for-N=4}, except that we now have $N=10$.} \label{Jx^2-plot-for-N=10}
\end{figure}

We now show that the effect of initial correlations are not manifested in the dynamics of $j_x$ alone, we show in Figs.~\ref{Jx^2-plot-for-N=4} and \ref{Jx^2-plot-for-N=10} the dynamics of $j_x^{(2)}\equiv 4\expval{J_x^2}/N^2$. Such an observable is relevant in the study of spin squeezing and entanglement. Once again, the initial correlations can affect the dynamics significantly.

\begin{figure}[htp] 
  \centering 
  \includegraphics[width=0.95\linewidth]{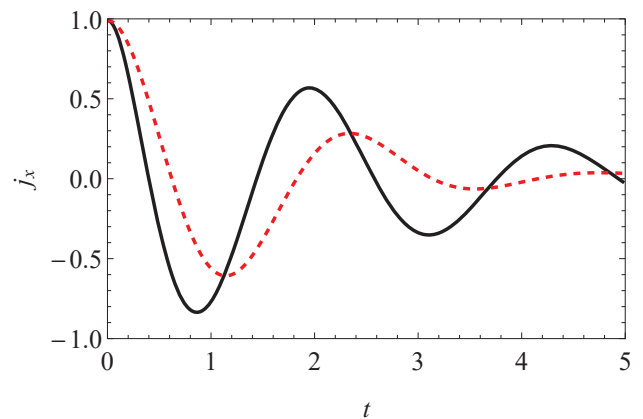}
\caption{Behavior of $j_x$ against $t$ for $N=4$ with (black, solid) and without (dashed, red) taking into
account initial correlations. Here we have used a sub-Ohmic environment with $s = 0.5$. We also have $\varepsilon_0 = 4$, $\varepsilon=2.5$ and $\Delta=\Delta_0=0.5$, while the rest of the parameters used are the same as Fig. \ref{Puredephasing-N=1-unitary}.} \label{subOhmic1}
\end{figure}

\begin{figure}[htp] 
  \centering 
  \includegraphics[width=0.95\linewidth]{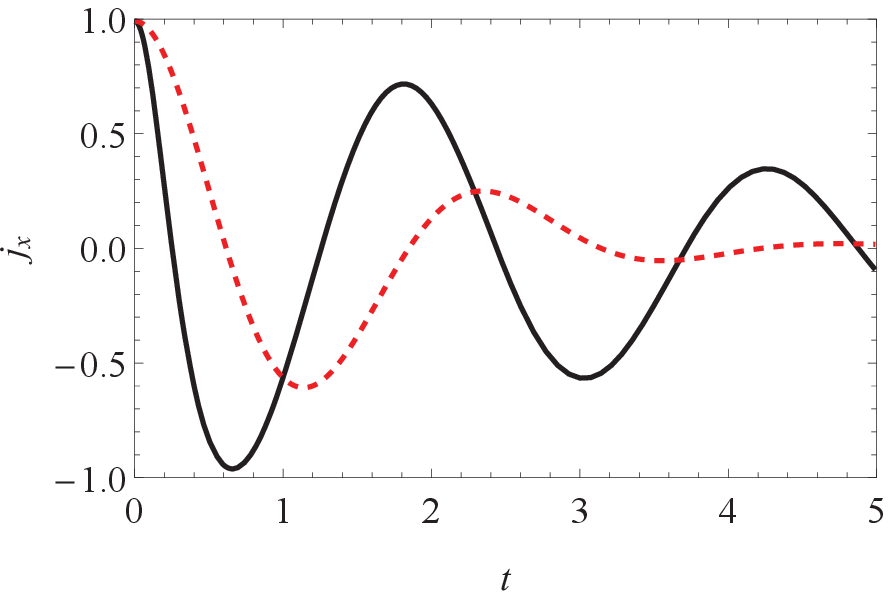}
\caption{Same as Fig.~\ref{subOhmic1}, except that now $N = 10$.} \label{subOhmic2}
\end{figure}

Finally, let us also demonstrate the effect of the initial correlations with sub-Ohmic environment, that is, $J(\omega) = G\omega^s \omega_c^{1 - s} e^{-\omega/\omega_c}$. Since these environments have longer correlation times, we expect that the effect of the initial correlations will be greater as well. This is indeed the case, as can be seen by comparing Figures \ref{subOhmic1} and \ref{subOhmic2} with Figs.~\ref{Beyond-PD-N=4} and Figs.~\ref{Beyond-PD-N=10} where an Ohmic environment had been used. 

\section{Application to the spin-spin environment model}

\begin{figure}[htp] 
  \centering 
  \includegraphics[width=0.95\linewidth]{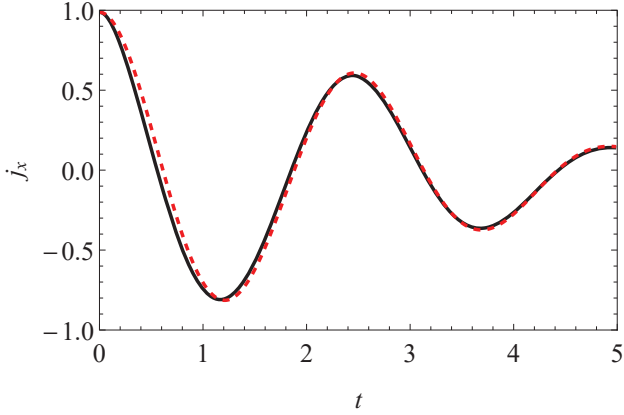}
\caption{Behavior of $j_x$ against $t$ for $N = 4$
with (black, solid) and without (dashed, red) taking into account initial correlations. Here we have $\varepsilon_0 = 4$, $\varepsilon=2.5$, $\Delta=\Delta_0=0.5$, $G = 0.05$, $\beta=1$ and $\omega_c=5$.} \label{spinenv1}
\end{figure}

\begin{figure}[htp] 
  \centering 
  \includegraphics[width=0.95\linewidth]{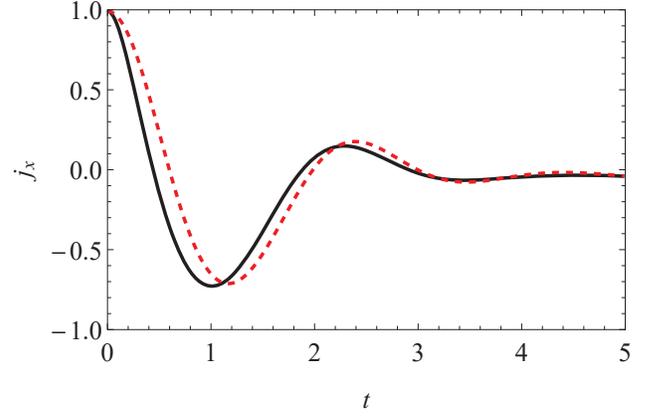}
\caption{Same as Fig.~\ref{spinenv1}, except that now $N = 10$.} \label{spinenv2}
\end{figure}

We consider now a collection of identical two-level systems interacting with an environment consisting of two-level systems \cite{cucchietti2005decoherence,CamaletPRB2007, Schlosshauerbook,VillarPhysLettA2009,SegalJCP2014}. We then have
\begin{align*}
    H_\text{S0}
    &= \varepsilon_0 J_z + \Delta_0 J_x\\
    H_S&= \varepsilon J_z + \Delta J_x\\
    H_B&= \sum_{k} \frac{\omega_k}{2} \sigma_{x}^{(k)},\\
    V&=J_x \otimes \sum_k g_k \sigma_{z}^{(k)}
\end{align*}
where $\sigma_{z}^{(k)}$ and $ \sigma_{x}^{(k)}$ are the Pauli $z$-spin and $x$-spin operators
of the $k$th environment spin respectively, $\omega_k$ denotes the tunneling matrix element for the $k^{\text{th}}$ environment spin and $g_k$ quantifies the coupling strength.\\
The different environment leads to a different correlation function $C_{ts}$ as well as a different factor $J^R_\text{corr}(\beta,t)$ that takes into account the effect of the initial system-environment correlations. We first note that   
\begin{align}
    B_{\text{corr}}(\lambda,t)&=\sum_k  |g_k|^2\Big\{\tanh{\left(\frac{\beta \omega_k}{2}\right)}e^{-\omega_{k}\left(\lambda-it\right)}\nonumber\\
    &+2n_k\sinh{\left(\lambda\omega_{k}-i\omega_{k}t\right)}\Big\},
\end{align}
with, 
\begin{align}
    n_k=\frac{1}{2}\left\{\coth{\left(\frac{\beta\omega_k}{2}\right)}-1\right\}\nonumber.
\end{align}
Since the factors $\alpha_1(\lambda,t)$, $\alpha_2(\lambda,t)$, and $\alpha_3(\lambda,t)$ are the same as before, this allows us to work out the role of the initial correlations [see Eq.~\eqref{initialcorrelations}]. In a similar manner, the environment correlation function can also be worked out. As before, to perform the sum over the environment modes, we
use the same procedure as mentioned earlier, that is, $\sum_k |g_k|^2 (...) \rightarrow \int_0^{\infty}d\omega J(\omega) (...)$. Results are shown in Figs.~\ref{spinenv1} and \ref{spinenv2}. Once again, the role of the initial correlations is relatively small for a smaller value of $N$. However, as $N$ increases, it is clear that we need to take into account the role of the initial correlations to obtain an accurate picture of the system dynamics.

\section{Conclusion}
To conclude, we have shown that if we start from the joint thermal equilibrium state of a quantum system and its environment, and then apply a unitary operation to the system to prepare the system quantum state, the initial correlations that exist in the joint thermal equilibrium state influence the subsequent dynamics of the system. We have derived a time-local master equation, correct to second-order in the system-environment coupling strength, that takes into account the effect of these correlations, showing therefore that one need not necessarily be in the strong system-environment coupling regime to observe the effects of the initial correlations. The structure of this master equation is very interesting, as the form of the term that takes into account the initial correlations is the same as the relaxation and dephasing term. In this sense, one can say that the initial correlations affect the decoherence and dephasing rates, a fact which has already been pointed out in studies of the role of initial correlations in pure dephasing models \cite{MorozovPRA2012}. Finally, we actually applied our master equation to the large spin-boson model as well as to a collection of two-level systems interacting with a spin environment to quantitatively investigate the role of the initial correlations. We found that when the number of spins is small, then the initial correlations do not play a significant role. However, for a larger number of spins, the initial correlations must be accounted for in order to explain the dynamics accurately.

\section*{acknowledgements}
A.~R.~M, M.~Z.~and A.~Z.~C. are grateful for support from HEC under grant No 5917/Punjab/NRPU/R\&D/HEC/2016. Support from the LUMS Faculty Initiative Fund is also acknowledged.

\appendix
\section{The relaxation/dephasing term in the master equation}
We look at the contribution of the third term in Eq.~\eqref{ref16}. We need to consider only $\rho_\text{tot}^{(0)}(0)$ will contribute since we are restricted to second-order terms in the master equation only. We obtain
\begin{align}
    &\alpha^2 \text{Tr}_\text{S,B}\left \{\rho_\text{tot}^{(0)}(0)\int_{0}^{t}\left[\left[\Tilde{V}(t),\Tilde{X}_{nm}(t)\right],\Tilde{V}(s)\right]ds\right \} \notag \\
    &=\alpha^2 \text{Tr}_\text{S,B}\left (\{\Tilde{\rho}_{S0}\otimes\rho_B) \int_{0}^{t}\left[\left[\Tilde{V}(t),\Tilde{X}_{nm}(t)\right],\Tilde{V}(s)\right]ds\right \} \notag \\
    &=\alpha^2 \text{Tr}_\text{S,B} \Bigg\{(\Tilde{\rho}_{S0}\otimes\rho_B)\Bigg(\int_{0}^{t}\Tilde{V}(t)\Tilde{X}_{nm}(t)\Tilde{V}(s)ds\nonumber\\
    &-\int_{0}^{t}\Tilde{V}(s)\Tilde{V}(t)\Tilde{X}_{nm}(t)ds-\int_{0}^{t}\Tilde{X}_{nm}(t)\Tilde{V}(t)\Tilde{V}(s)ds\nonumber\\
    &+\int_{0}^{t}\Tilde{V}(s)\Tilde{X}_{nm}(t)\Tilde{V}(t)ds\Bigg) \Bigg\}.
\end{align}  
This splits up into four terms. The first term is 
\begin{align*}    
    &\alpha^2 \text{Tr}_\text{S,B}\left \{(\Tilde{\rho}_{S0}\otimes\rho_B)\int_{0}^{t}\Tilde{V}(t)\Tilde{X}_{nm}(t)\Tilde{V}(s)ds\right \} \\
    &=\alpha^2\int_{0}^{t}\text{Tr}_\text{S,B}\Big \{\Tilde{\rho}_{S0}\otimes\rho_BU_0^\dagger(t)VU_0(t)U_0^\dagger(t)X_{nm}U_0(t)\\\
    &\times U_0^\dagger(s)VU_0(s)\Big \}\,ds\\
    &=\alpha^2\int_{0}^{t}\text{Tr}_\text{S}\left \{\Tilde{\rho}_{S0}U_S^\dagger(t)FY_{nm}U_S(t,s)FU_S(s)\right \}\\\
    &\times \text{Tr}_\text{B}\left \{\rho_BB(t)B(s)\right \}ds\\
    &=\alpha^2\int_{0}^{t}\bra{m}\Bar{F}(t,s)\Tilde{\rho}_{S0}(t)F\ket{n} C_{ts} ds.
\end{align*}
In a similar way, we can simplify the other terms. Putting them all back together, and shifting to the basis-independent representation, we obtain the third term in Eq.~\eqref{finalme}.

\section{Pure dephasing model}
\label{appendixsinglespingate}
For completeness, we now sketch the derivation of the system dynamics for the large spin-boson model. We have $N$ identical two-level systems interacting with a common environment of harmonic oscillators. The dynamics of such a system can be described by the Hamiltonian
\begin{align*}
    H_\text{tot}=H_S+H_B+V, \nonumber
\end{align*}
where $H_S = \varepsilon J_z$, $H_B= \sum_k \omega_k b_{k}^{\dagger} b_k$, and $V=J_z \sum_{k} \left(g_{k}^{*}b_k + g_k b_k^\dagger \right)$. In the interaction picture, the Hamiltonian becomes 
\begin{align*}
    H_I \left(t\right)&=e^{i \left(H_B+H_S\right) t} V  e^{-i  \left(H_B+H_S\right)t}\nonumber\\ 
    &=J_z \sum_{k} \left(g_{k}^{*}b_k e^{-i\omega_k t} + g_k b_k^\dagger e^{i\omega_k t}\right).
\end{align*}
The corresponding unitary time-evolution operator can be found exactly via the Magnus expansion. Consequently, the unitary operator for the complete Hamiltonian is 
\begin{align*}
    U\left( t \right) = e^{-iH_S t} e^{-i H_B t} U_I\left( t \right) \nonumber \\
     U\left( t \right) = e^{-i\varepsilon J_z t} e^{-i H_B t} U_I\left( t \right),
\end{align*}
where,
\begin{align*}
    U_I\left( t \right)= \text{exp} \left\{ J_z \sum_k \left[ \alpha_k \left(t\right)b_k^\dagger -  \alpha_k^* \left(t\right)b_k \right] -iJ_z^2 t\Delta \left( t \right)  \right\}.
\end{align*}
Here $\alpha_k\left( t \right)
=  \frac{g_k\left(1-e^{-i\omega_k t} \right)}{\omega_k}$ and $\Delta \left( t \right) = \frac{1}{t} \sum_k \abs{g_k}^2  \frac{\left[\sin(\omega_k t) - \omega_k t\right]}{\omega_k^2}$. The reduced density operator of the system is then written in the $J_z$ eigenbasis as 
\begin{align}
    \left[\rho \left(t\right)\right]_{mn} = \text{Tr}_{S,B} \left[ U\left(t\right) \rho \left(0\right)U^\dagger\left(t\right) P_{nm} \right] \label{eqref2}
\end{align}
where $P_{nm} \equiv \ket{n}\bra{m},$ such that $J_z\ket{n}=n\ket{n}.$ We can write Eq.~\eqref{eqref2} in the Heisenberg picture, where $P_{nm}\left(t\right) = U^\dagger \left( t \right) P_{nm} U\left(t\right)$ is the Heisenberg picture operator. It follows that
\begin{align*}
    \left[\rho \left(t\right)\right]_{mn} = \text{Tr}_{S,B} \left[ \rho \left(0\right) P_{nm}\left(t\right) \right].
\end{align*}
It is straightforward to find that 
\begin{align*}
    P_{nm}\left(t\right) = e^{-i\varepsilon \left( m - n \right) t} e^{-i\Delta \left( t \right) \left( m^2 - n^2 \right) t} e^{-R_{nm}\left(t\right) }P_{nm},
\end{align*}
where,
\begin{align*}
    R_{nm}\left(t\right) = \left(n-m\right) \sum_k \left[ \alpha_k \left(t\right)b_k^\dagger -  \alpha_k^* \left(t\right)b_k \right].
\end{align*}
Hence, it follows that,
\begin{align}
    \left[\rho \left(t\right)\right]_{mn} =& e^{-i\varepsilon \left( m - n \right) t} e^{-i\Delta \left( t \right) \left( m^2 - n^2 \right) t} \times \nonumber  \\ & \text{Tr}_{S,B} \left[ \rho \left(0\right) e^{-R_{nm}\left(t\right)} P_{nm} \right]\label{ref3} 
\end{align}
This is a general result because we have not yet defined the joint system-environment initial state $\rho(0)$. Hence, it can be applied to both uncorrelated and correlated initial states. We will derive an expression for $\left[\rho_S \left(t\right)\right]_{mn} $ for both cases. First, without considering the initial system-environment correlations, 
\begin{align*}
    \rho_{\text{tot}} \left(0\right) = \rho \left(0\right) \otimes \rho_B
\end{align*}
where, $\rho_B = \frac{e^{-\beta H_B}}{Z_B} \  \text{with,}  \  Z_B = \text{Tr}_B \left[ e^{-\beta H_B} \right]$. We then get 
\begin{align}
    \left[\rho \left(t\right)\right]_{mn} =& \left[\rho \left(0\right)\right]_{mn}e^{-i\varepsilon \left( m - n \right) t} e^{-i\Delta \left( t \right) \left( m^2 - n^2 \right) t} \times \nonumber\\  & \text{Tr}_{B} \left[ \rho_Be^{-R_{nm}\left(t\right)}\right]. \label{ref4}
\end{align}
Simplifying the trace over the environment, we obtain
\begin{align}
    \left[\rho \left(t\right)\right]_{mn} =& \left[\rho \left(0\right)\right]_{mn}e^{-i\varepsilon \left( m - n \right) t} e^{-i\Delta \left( t \right) \left( m^2 - n^2 \right) t} \times \nonumber \\ & e^{-\gamma\left(t\right)\left( m-n\right)^2t},
\end{align}
with
\begin{align}
    \gamma\left(t\right)
    =\frac{1}{t}\sum_k   \frac{4\abs{g_k}^2\left[1-\cos\left(\omega_k t \right)\right]}{\omega_k^2} \coth{\left( \frac{\beta \omega_k}{2}\right)}.
\end{align}
Next, we look at the case where the initial system-environment state is of the form 
\begin{align}
\rho \left(0\right) = \frac{\Omega e^{-\beta H_\text{tot}} \Omega^\dagger}{Z}, \label{ref5}
\end{align}
where $Z= \text{Tr}_{S,B}\left[\Omega e^{-\beta H_\text{tot}} \Omega^\dagger\right], $ while  $\Omega$ is a unitary operator. We simplify $Z$ first by introducing a completeness relation, 
\begin{align*}
    Z&= \sum_l \text{Tr}_{S,B}\left[\Omega e^{-\beta H_\text{tot}}\ket{l}\bra{l} \Omega^\dagger\right], \nonumber \\
    &=\sum_l e^{-\beta \varepsilon l}\bra{l}\Omega^\dagger \Omega \ket{l} \text{Tr}_{B}\left[ e^{-\beta H_B^{\left(l\right)}} \right],
\end{align*}
where, 
\begin{align*}
    H_B^{\left(l\right)} = H_B + l \sum_k \left( g_{k}^{*}b_k + g_k b_k^\dagger \right).
\end{align*}
Using the displaced harmonic oscillator modes\\ $B_{k,l}= b_k + \frac{lg_k}{\omega_k}$ and $B_{k,l}^\dagger= b_k^\dagger + \frac{lg_k^*}{\omega_k}$, we obtain
\begin{align*}
    Z= \sum_l e^{-\beta \varepsilon l}\bra{l}\Omega^\dagger \Omega \ket{l} e^{\beta l^2 \mathcal{C}} Z_B,
\end{align*}
where, $\mathcal{C}= \sum_k \frac{4 \abs{g_k}^2}{\omega_k}$. Proceeding further, again using a completeness relation, we note that 
\begin{align*}
    \left[\rho \left(t\right)\right]_{mn} &=\frac{1}{Z} \sum_ l e^{-i\varepsilon \left( m - n \right) t} e^{-i\Delta \left( t \right) \left( m^2 - n^2 \right) t} \times \nonumber \\ & \bra{l}\Omega^\dagger P_{nm}\Omega \ket{l}  \text{Tr}_{B} \left[ {e^{-\beta H_B^{\left(l\right)}}} e^{-R_{nm}\left(t\right)}\right] \label{ref6},
\end{align*}
writing $R_{nm}\left(t\right)$ as 
\begin{align*}
     R_{nm}\left(t\right) = \left(n-m\right) \sum_k \left[ \alpha_k \left(t\right)b_k^\dagger -  \alpha_k^* \left(t\right)b_k \right] + i \Phi_{nm}^{\left(l\right)}\left(t\right),
\end{align*}
where,
\begin{align*}
\Phi_{nm}^{\left(l\right)}\left(t\right)
&= 2l\left(n-m\right) \Phi\left(t\right),\\ \Phi\left(t\right) 
&= \sum_k \frac{4\abs{g_k}^2}{\omega_k^2}\sin\left(\omega_k t\right).
\end{align*}
It can be shown that
\begin{align*}
    \text{Tr}_{B} \left[ {e^{-\beta H_B^{\left(l\right)}}} e^{-R_{nm}\left(t\right)}\right] = e^{-i\Phi_{nm}^{\left(l\right)}\left(t\right)}e^{\beta l^2 \mathcal{C}} Z_B e^{-\gamma\left(t\right)\left( m-n\right)^2 t}.
\end{align*}
Hence, we obtain
\begin{align}
    &\left[\rho \left(t\right)\right]_{mn} =\left[\rho \left(0\right)\right]_{mn} e^{-i\varepsilon \left( m - n \right) t} e^{-i\Delta \left( t \right) \left( m^2 - n^2 \right) t}\times \nonumber \\ & \, \, \, \,\, \, \, \,\, e^{-\gamma\left(t\right)\left( m-n\right)^2 t}  \frac{\sum_l \left[\bra{l}\Omega^\dagger P_{nm}\Omega \ket{l} e^{-i\Phi_{nm}^{\left(l\right)}\left(t\right)}e^{-\beta \varepsilon l}e^{\beta l^2 \mathcal{C}}\right]}{\sum_l \left[\bra{l}\Omega^\dagger P_{nm}\Omega \ket{l} e^{-\beta \varepsilon l}e^{\beta l^2 \mathcal{C}}\right]}.
\end{align}

\end{document}